\documentclass[prl,amsmath,amssymb,twocolumn, longbibliography]{revtex4-1}

\usepackage{graphicx}
\usepackage{subfigure}
\usepackage{adjustbox}
\usepackage{bm}
\usepackage{color}
\usepackage{braket}
\usepackage{standalone}
\usepackage{multirow}
\usepackage{tikz}
\usepackage{dsfont}
\usepackage[colorlinks,bookmarks=true,citecolor=blue,linkcolor=red,urlcolor=blue]{hyperref}
\newcommand{\bea}{\begin{eqnarray}}
\newcommand{\eea}{\end{eqnarray}}
\definecolor{dgreen}{rgb}{0.1,0.5,0.1}
\definecolor{babyblue}{rgb}{0.54, 0.81, 0.94}

\begin{document}

\title{Dynamically Induced Exceptional Phases in Quenched Interacting Semimetals}
\author{Carl Lehmann$^1$}
\author{Michael Sch\"uler$^2$}
\author{Jan Carl Budich$^1$}
\email{jan.budich@tu-dresden.de}
\affiliation{$^1$Institute of Theoretical Physics${\rm ,}$ Technische Universit\"{a}t Dresden and W\"{u}rzburg-Dresden Cluster of Excellence ct.qmat${\rm ,}$ 01062 Dresden${\rm ,}$ Germany}
\affiliation{$^2$Stanford Institute for Materials and Energy Sciences (SIMES)${\rm ,}$SLAC National Accelerator Laboratory${\rm ,}$Menlo Park${\rm ,}$CA 94025${\rm ,}$USA}
\date{\today}

\begin{abstract}
We report on the dynamical formation of exceptional degeneracies in basic correlation functions of non-integrable one- and two-dimensional systems quenched to the vicinity of a critical point. Remarkably, fine-tuned semi-metallic points in the phase diagram of the considered systems are thereby promoted to topologically robust non-Hermitian (NH) nodal phases emerging in the coherent long-time evolution of a dynamically equilibrating system. In the framework of non-equilibrium Green's function methods within the conserving second Born approximation, we predict observable signatures of these novel NH nodal phases in simple spectral functions as well as in the time-evolution of momentum distribution functions.  

\end{abstract}

\maketitle

In the realm of thermal equilibrium, the basic understanding of (semi-)metallic phases is largely based on the notion of quasi-particles, the life-time of which diverges at the Fermi surface when approaching zero temperature \cite{Landau1957,AshcroftMermin}. At finite temperature, inter-particle scattering in correlated systems generically induces a finite life-time of elementary excitations as described by a non-Hermitian (NH) self-energy $\Sigma$. 

Interestingly, if $\Sigma$ acquires a complex matrix-structure in some internal degrees of freedom, topologically stable exceptional points (EPs) \cite{Berry2004,Heiss2012,Miri2019} at which basic correlation functions of the system become non-diagonalizable may arise \cite{Kozii2017,Bergholtz2019,Yoshida2020,Rausch2021}. Such genuinely NH spectral properties of Green's functions are hallmarked by open Fermi arcs connecting the EPs along with a characteristic non-analytical dispersion $\sim \sqrt{q}$ emanating from each EP, where $q$ denotes (lattice) momentum relative to the EP. In a broader context, this behavior exemplifies the notion of NH topological phases that have become a broad frontier of interdisciplinary research in recent years \cite{Rudner2009,Zeuner2015,Malzard2015,Lee2016,Zhou2018,Lieu2018,Longhi2018,Kunst2018,Yao2018,Imhof2018,Gong2018,Kawabata2019_1,Budich2019,Yoshida2019,Cerjan2019,Kawabata2019_2,Kunst2019,Song2019,Xiao2020,Weidemann2020,Budich2020}  (see Ref.~\cite{review} for an overview). 

\begin{figure}[htp]
    \centering
    \includegraphics[width=1.0\linewidth]{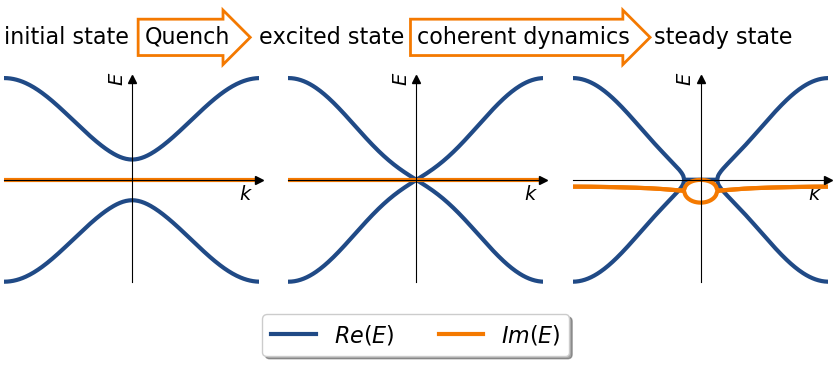}
    \caption{Illustration of the quench protocol and the complex energy bands of the effective NH Hamiltonian (cf.~Eq.~(\ref{eqn:heff})) around the Fermi energy. As inital state, we consider the ground state of a band-insulator at half filling (left). The system Hamiltonian then undergoes a parameter-quench in which the band gap is closed thus leading to a critical non-equilibrium state (middle), and thereafter an interaction between the particles is switched on so as to make the Hamiltonian non-integrable. During the subsequent coherent post-quench dynamics, the excitations created during the quench dynamically equilibrate. In this process, a nodal NH phase featuring exceptional degeneracies (hallmarked by a degeneracy of both the real and the imaginary part) emerges (right). }
    \label{fig1}
\end{figure}

Here, we demonstrate how such novel exceptional NH phases may dynamically emerge in the coherent time-evolution of non-integrable fermionic  systems that are prepared in an insulating {\em zero-temperature} state before undergoing a quantum quench to the vicinity of a critical point (see Fig.~\ref{fig1} for an illustration). To analyze the quasi-particle excitations long time after the quench, we study the effective NH Hamiltonian

\begin{align}
H_{\mathrm{NH}}(k,\omega) = H_f(k) +\Sigma^R(k,\omega), 
\label{eqn:heff}
\end{align}

where $H_f(k)$ denotes the non-interacting part of the (Hermitian) post-quench Hamiltonian in reciprocal space, and $\Sigma^R(k,\omega)$ is the Fourier-transform of the late-time retarded self-energy. In both one-dimensional (1D) and two-dimensional (2D) systems, we find that fine-tuned critical points in the post-quench Hamiltonian are dynamically promoted to topologically stable NH nodal phases of $H_{\mathrm{NH}}$ (see Eq.~(\ref{eqn:heff})). More precisely, isolated EPs occurring in the $k$-$\omega$ plane of 1D systems are pinned to the Fermi energy ($\omega =0$) only in the presence of a pseudo-Hermiticity constraint such as chiral symmetry, while in 2D EPs at the Fermi-energy are generically stable against small perturbations of the post-quench band-structure. We identify clear observable signatures of the predicted non-equilibrium phase-diagrams in basic spectral functions as well as in the transient behavior of momentum distribution functions during the dynamical equilibration. Our general findings are quantitatively corroborated by numerical simulations of fully microscopic lattice models in both 1D and 2D within the framework of the non-equilibrium Green's function (NEGF) approach in conserving second Born approximation (2BA) \cite{Stefanucci_book}.

{\it Quench protocol and computational implementation.---}
We study the coherent time-evolution of fermionic systems in both 1D and 2D that are initially prepared in the ground-state of a band-insulator Hamiltonian $H_i$, before undergoing a parameter quench $H_i \rightarrow H_f$ to a final Hamiltonian $H_f$ in two stages: first, the band-structure parameters are tuned to the vicinity of a critical point, and, second, interactions between the particles are switched on so as to make the system {\em non-integrable}. During the quench, a finite density of excitations is created thus bringing the system far from thermal equilibrium. However, due to the fully chaotic nature of $H_f$, the system may dynamically equilibrate and approach a steady-state in the long-time limit. In this process, the scattering between the particles effectively represents a source of dissipation that may drive the elementary excitations described by the effective Hamiltonian $H_{\mathrm{NH}}(k,\omega)$ (see Eq.~\ref{eqn:heff})  into intriguing non-Hermitian topological phases, as we demonstrate below. The considered quench protocol is illustrated in Fig.~\ref{fig1}.

To compute the time-evolution of a system over this quench protocol, we solve the Kadanoff-Baym equations (KBEs) for the Green's function \cite{Stefanucci_book,balzer_nonequilibrium_2012,aoki_nonequilibrium_2014,schlunzen_nonequilibrium_2016,nessi2020}, using the software package NESSi \cite{nessi2020}. Long time after the quench, where the Hamiltonian stays constant in time, the retarded Green's functions may only depend on the time difference, i.e. $G^R(k,t + \Delta t, t) \rightarrow G^R(k,\Delta t)$. As a consequence, we may Fourier transform to frequency space as $G^R(k,\omega) = \int_{-\infty}^{\infty} d(\Delta t)  G^R(k,\Delta t) e^{i\omega \Delta t}$. In this regime, the effective Hamiltonian $H_{\text{NH}}(k,\omega)$ (see Eq.~({\ref{eqn:heff}})) then results from the Dyson equation in frequency space as
\begin{align}
	& G^R(k,\omega) \left(\omega - H_{\text{NH}}(k,\omega)\right) = \mathds{1} \label{Kadanoff_Baym}.
\end{align}

We emphasize that we do by no means assume the formation of a steady-state in our simulations. Instead, we simulate the full non-equilibrium time-evolution of the considered systems over the quench, and identify from the resulting data a late-time regime, where $G^R$ may to a good approximation be described by Eq.~(\ref{Kadanoff_Baym}).~\footnote{We calculated the effective Hamiltonian $H_{\text{NH}}(k,\omega)$ by Eq.~(\ref{Kadanoff_Baym}) with the retarded Green's function and by Eq.~({\ref{eqn:heff}}) with the retarded self-energy $\Sigma^R(k,\omega)$. Both ways lead to the same results, in this paper we used the latter method, because $\Sigma^R(k,\Delta t)$ decays much faster in the late time regime, thus Fourier transformation was numerically easier.} By contrast to the retarded Green's function, the momentum distribution functions encoded in the lesser component of the Green's function are found to exhibit an interesting transient behavior, where the exceptional NH phase is reflected in a striking slow-down of thermalization. 

{\it Dynamically induced exceptional quasi-particles.---}
For concreteness and simplicity, from now on we focus on two-banded systems, noting that the generalization of our discussion to a higher number of bands will be straightforward. The effective Hamiltonian  $H_{\text{NH}}(k,\omega)$ (cf.~Eq.~(\ref{eqn:heff})) then is a $2\times2$ matrix that may be written as
\begin{align}
	H_{\text{NH}}(k,\omega) = d_0 \sigma_0 + (\mathbf{d}_{\text{R}}+i\mathbf{d}_{\text{I}}) \cdot \bm{\sigma},
\end{align}
where $\bm{\sigma}$ denotes the vector of standard Pauli matrices, the identity matrix is denoted by $\sigma_0$, and the dependence of $\mathbf{d}_{\text{R}}, \mathbf{d}_{\text{I}} \in \mathbb R^3,~d_0 \in \mathbb C$ on momentum and frequency has been suppressed for brevity of notation. The corresponding complex energy eigenvalues are given by  $E_{\pm}= d_0 \pm \sqrt{\mathbf{d}_{\text{R}}^2 - \mathbf{d}_{\text{I}}^2  + 2i\mathbf{d}_{\text{R}} \cdot \mathbf{d}_{\text{I}} }$. Degeneracies in the spectrum generically occur in the form of exceptional (i.e. non-diagonalizable) points as non-trivial solutions to the equations

\begin{align}
&\mathbf{d}^2_{\text{R}} - \mathbf{d}^2_{\text{I}} = 0,\label{cond_1}\\
&\mathbf{d}_{\text{R}} \cdot\mathbf{d}_{\text{I}} = 0. \label{cond_2}
\end{align}	

Simple parameter counting shows that EPs in 1D occur at isolated points in the  $k-\omega$ plane, namely at the intersections of the contours satisfying the individual conditions (\ref{cond_1}) and (\ref{cond_2}), respectively. Thus, to stabilize EPs at the Fermi-energy (i.e. at $\omega =0$) in 1D, an additional symmetry such as the chiral symmetry 
\begin{align}
	\sigma_z H_{\text{NH}}^{\dagger} (k,\omega)\sigma_z = - H_{\text{NH}}(k,-\omega), \label{eqn:cs}
\end{align}

that reduces the co-dimension of EPs at $\omega =0$ by trivializing (\ref{cond_2})  is required \cite{Budich2019,Yoshida2019}. By contrast, in the three-dimensional $k-\omega$ volume of a 2D system, (\ref{cond_1}) and (\ref{cond_2}) are simultaneously fulfilled along exceptional lines that may cross the Fermi-energy at isolated EPs. Thus, EPs in 2D are topologically stable independent of symmetries. Generally,  the EPs occur in pairs that are connected by so-called Fermi arcs (mathematically branch cuts) at which (\ref{cond_2}) and $\mathbf{d}^2_{\text{R}} - \mathbf{d}^2_{\text{I}} < 0$  are satisfied, thus giving rise to a continued degeneracy in the real part of the spectrum.

From the above general analysis, the concrete mechanism by which $H_{\text{NH}}(k,\omega=0)$ is found to develop EPs in the considered physical setup may be understood as follows. Since the non-interacting part $H_f(k)$ of the post-quench Hamiltonian is assumed to be at (least in the vicinity of) a critical phase exhibiting gap-closing points, $\lvert \mathbf{d}_{\text{R}} (k, \omega=0)\rvert$ is very small in those regions in reciprocal space. At the same time, excitations are created during the quench predominantly in the same near-critical regions, thus inducing a sizable NH self-energy (leading to finite $\lvert \mathbf{d}_{\text{I}}\rvert$) during the dynamical equilibration process. As a consequence, in 1D Eq.~(\ref{cond_1}) is satisfied at pairs of points enclosing the (almost) band-touching points of $H_f(k)$, and in 2D Eq.~(\ref{cond_1}) is satisfied on contours around those points in momentum space, respectively. In 1D, the chiral symmetry (\ref{eqn:cs}) then stabilizes EPs by assuring that also Eq.~(\ref{cond_2}) is satisfied, while in 2D Eq.~(\ref{cond_2}) is generically satisfied at isolated EPs on the aforementioned contour encircling the band-touching point of $H_f(k)$ \cite{review}. In both cases, basic continuity arguments yield that no fine-tuning of $H_f(k)$ to an exact critical phase is required. Instead, the value of $\lvert \mathbf{d}_{\text{I}}\rvert$ in the near-critical region that sets the scale for the robustness of the EPs is determined by the density of excitations created during the quench as well es the residual energy gap.

{\it Microscopic 1D model and numerical analysis.---}
We consider an extended Su-Schrieffer-Heeger (SSH) model \cite{Su1979,Kruckenhauser2018} described by the tight-binding Hamiltonian
\begin{align}
H_{\text{SSH}} (t) = - \frac{1}{2} \sum_{j} &\left[ J(t) \mathbf{\Psi}_j^{\dagger}  \sigma_x \mathbf{\Psi}_j +  
\mathbf{\Psi}_j^{\dagger} \left(\tau \sigma_x -i d \sigma_y \right) \mathbf{\Psi}_{j+1} \right] \nonumber \\
& + \text{h.c.},
\label{Hssh}
\end{align}
where $\mathbf{\Psi}_j^{\dagger} = \left(a_{j}^\dagger, b_{j}^\dagger \right)$ denotes the spinor in A-B sublattice space of creation operators acting on unit cell $j$, and $J(t),d, \tau$ are real hopping parameters.
To this free model we add the density-density interaction
\begin{align}
H_{\text{I}}(t) &= \sum_{<i,j>} \sum_{s = \left\lbrace a,b \right\rbrace} U_{s}(t) \left(n_i^{s} - \frac{1}{2}\right) \left(n_{j}^{s} - \frac{1}{2}\right), 
\label{Hint}
\end{align}
where $n_j^{a}=a_{j}^{\dagger} a_{j}$ ($n_j^{b} = b_{j}^{\dagger} b_{j}$) are the density operators corresponding to the sublattice sites $s = \left\lbrace a,b \right\rbrace$, the sum on $<i,j>$ is over nearest neighbor unit cells, and $U_a(t), U_b(t)$ are the interaction parameters.
We note that the total Hamiltonian $H= H_{\text{SSH}} + H_{\text{I}}$ at the considered half-filling obeys chiral symmetry (cf.~Eq.~(\ref{eqn:cs})) due to its invariance under the transformations $a_j \rightarrow a_{j}^{\dagger}$, $b_{j} \rightarrow - b_{j}^{\dagger}$, $i \rightarrow -i$ \cite{Chiu2016}. Assuming periodic boundary conditions, the Bloch Hamiltonian associated with $H_{\text{SSH}}$ reads as
\begin{align}
H_{\text{SSH}} (k,t) &=  \mathbf{d}_{\text{SSH}}(k,t) \cdot \bm{\sigma},
\label{sshbloch}
\end{align}
where $\mathbf{d}_{\text{SSH}}(k,t) = \left(-J(t) - \tau \cos(k), - d \sin(k), 0 \right)$ is the Bloch vector. To implement the above quench protocol, we vary with time  $J$ from a gapped regime $\lvert J(t_i)\rvert > \lvert \tau \rvert$ to the critical point at $J(t_f) = \tau$. Thereafter, we switch on the interaction, where $U_a \ne U_b$ is important to obtain a self-energy with a non-trivial matrix-structure in A-B sublattice space \cite{Rausch2021}.

\begin{figure}[t]
	\centering
	\includegraphics[width=1.0\linewidth]{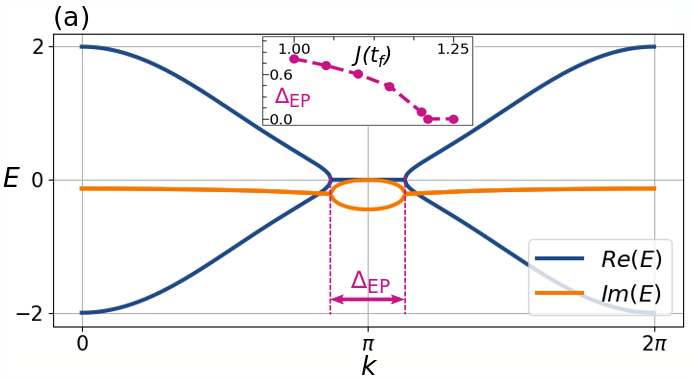}
	\includegraphics[width=1.0\linewidth]{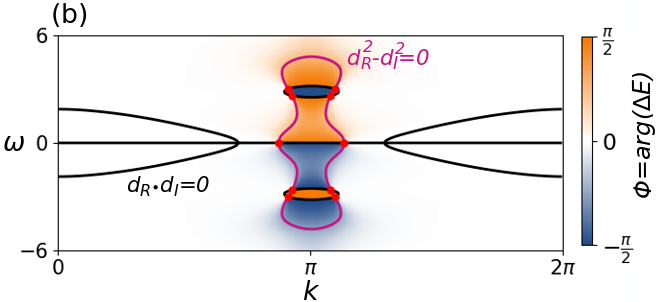}
	\caption{(a) Complex band-structure of the effective Hamiltonian $H_{\mathrm{NH}}(k,\omega =0)$ (cf.~Eq.~(\ref{eqn:heff})) at the Fermi energy in the long-time limit. (b) Visualization of the phase $\phi = \text{arg} \left( \sqrt{\mathbf{d}_{\text{R}}^2 - \mathbf{d}_{\text{I}}^2  + 2i\mathbf{d}_{\text{R}} \cdot \mathbf{d}_{\text{I}} } \right)$ of the complex energy gap function $\Delta E$. Contours at which conditions (\ref{cond_1}) and (\ref{cond_2}) are satisfied are indicated in violet and black, respectively. EPs (red dots) occur at the intersections of those contours. All EPs are connected pairwise by Fermi arcs along which (\ref{cond_2}) is satisfied and where $\mathbf{d}^2_{\text{R}} - \mathbf{d}^2_{\text{I}} < 0$. When crossing a Fermi arc, the phase $\phi$ jumps by $\pi$. Parameters in all plots are: $J(t_i)=-2.0, U_a=0.0 \rightarrow J(t_f)=1.0, U_a=1.5$ over the quench, all other parameter constant in time at $U_b=0.0$, $d=0.5$, $\tau=1.0$. System size $L=2\times250$ (i.e. $250$ particles and unit cells).
	Inset of (a): Stability of the NH exceptional phase against opening a small gap in $H_f(k)$ by detuning $J(t_f)$ from its critical value $1.0$, as determined by the length of the Fermi arc $\Delta_{\text{EP}}$ (system size $L=2\times80$). }
	\label{fig_1D_1}
\end{figure}

In Fig. \ref{fig_1D_1}, we present data on the occurrence of an exceptional NH phases in the dynamically equilibrated effective Hamiltonian. In particular, we microscopically corroborate the qualitative picture of the exceptional NH phase shown in Fig.~\ref{fig1} by computing the complex band structure of the effective Hamiltonian $H_{\mathrm{NH}}(k,\omega =0)$ (cf.~Eq.~(\ref{eqn:heff})) at the Fermi energy (see Fig. \ref{fig_1D_1}(a)). In Fig.~\ref{fig_1D_1}(b), we visualize the  phase  $ \phi (k,\omega)  = \text{arg} \left( \Delta E (k,\omega) \right)$ of the complex energy gap function $\Delta E = 2\sqrt{\mathbf{d}_{\text{R}}^2 - \mathbf{d}_{\text{I}}^2  + 2i\mathbf{d}_{\text{R}} \cdot \mathbf{d}_{\text{I}} }$ of $H_{\mathrm{NH}}(k,\omega)$. The contours in $k-\omega$ space at which conditions (\ref{cond_1}) and (\ref{cond_2}) are individually satisfied are found to intersect at two EPs right at the Fermi energy ($\omega =0$). 
Those EPs are pinned to $\omega=0$ by chiral symmetry (\ref{eqn:cs}), and can only disappear by local recombination in momentum space, i.e. by contracting the Fermi arc connecting them. Clearly, $ \phi$  exhibits a discontinuity in the form of a jump by $\pi$ at the Fermi arcs connecting the EPs, thus confirming their mathematical nature as branch cuts of $\Delta E$. In addition to those at the Fermi-energy, further EPs at finite frequencies are found to occur. Furthermore, we varied the post-quench parameter $J(t_f)$ so as to detune the post-quench Hamiltonian from the critical point. 
We find that the dynamically induced EPs are no fine-tuned phenomenon, but are robust against such small parameter changes. Specifically, on opening a small gap in the free post-quench Hamiltonian $H_f(k)$, the length $\Delta \textrm{EP}$ of the Fermi arc continuously shrinks, eventually leading to a recombination of the EPs at a sizable detuning of $J(t_f)$ from the critical point (see inset of Fig.~\ref{fig_1D_1}(a)).

\begin{figure}[t]
	\centering
	\includegraphics[width=1.0\linewidth]{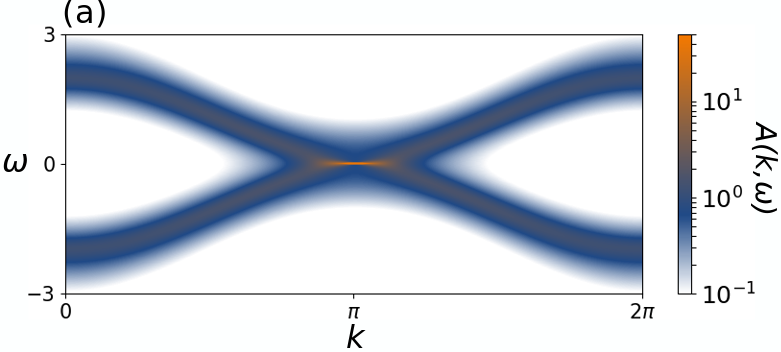}
	\includegraphics[width=1.0\linewidth]{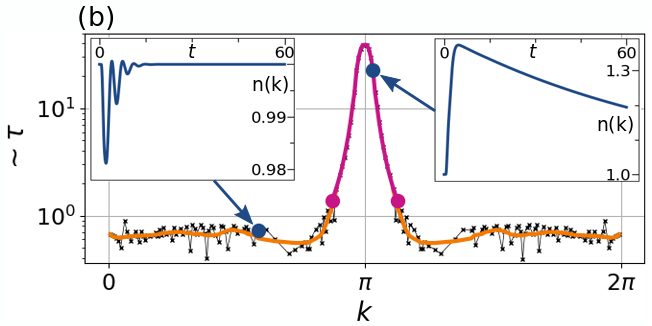}
	\caption{\label{fig3} (a) Spectral function $A(k,\omega) = - \dfrac{1}{\pi} {\text{Im}} \left( {\text{Tr}} \left( G^R(k,\omega) \right) \right)$ in 1D. Besides energy bands with moderate intensity, the Fermi arc characteristic of the exceptional NH phase is visible as strip with high intensity around momentum $k=\pi$ which slightly broadens close to the EPs. (b)  Decay times $\tau(k)$ for the decay of the non-equilibrium occupations $n(k,t)$. The decay times (black crosses) exhibit fluctuations, but follow a characteristic local mean (solid line): Within the Fermi arc (violet part, violet dots represent the EPs) the decay times are striking higher than outside of it (orange part). Insets: Real-time evolution of the decay of $n(k,t)$, distinguishing the Fermi arc region from the outer region. Parameters in all plots identical to Fig.~\ref{fig_1D_1}).}
\end{figure}

{\it Experimental signatures.---} To predict observable fingerprints of the dynamically induced exceptional NH phase, we compute the spectral function $A(k,\omega) = - \dfrac{1}{\pi} \text{Im} \left( \text{Tr} \left( G^R(k,\omega) \right) \right)$ associated with the effective Hamiltonian $H_{\text{NH}}(k,\omega)$ via Eq.~(\ref{Kadanoff_Baym}). Our numerical results are shown in Fig.~\ref{fig3}(a). At the Fermi energy, the extended Fermi arc connecting the EPs is clearly visible, thus allowing a distinction between the exceptional NH phase and ordinary nodal band-structure in basic spectroscopic measurements.

In addition, we study the time-dependence of the momentum distribution functions of the occupation number $n(k,t)$. During the dynamical equilibration process, $n(k,t)$ may obey fluctuations in time which relax in the long time limit towards a thermal value corresponding to a Gibbs ensemble $\rho =e^{-\beta H(t_f)} \slash \text{Tr} \left( e^{-\beta H(t_f)} \right)$ with respect to the post-quench Hamiltonian. There, the effective inverse temperature $\beta$ is determined by requiring that the total energy expectation value concurs with that of the time-evolved state after the quench, as determined by the Galitskii-Migdal formula \cite{Stefanucci_book}. Interestingly, we observe a striking difference in the thermalization process between the retarded Green's function and the momentum distributions. While $G^R$ quite rapidly approaches a steady state, thus justifying the use of Eq.~(\ref{Kadanoff_Baym}), we find that $n(k,t)$ exhibits a rich transient behavior that clearly hallmarks the dynamical formation of the exceptional NH phase (see Fig.~\ref{fig3}(b)). In particular, along the Fermi arc, the equilibration of  $n(k,t)$ is remarkably slow and follows an exponential decay reminiscent of an overdamped oscillator. By contrast, outside of the Fermi arc the occupation exhibits damped oscillations that decay rapidly. To quantify this behavior, in Fig.~\ref{fig3}(b) we show a simple numerical estimate of the momentum-dependent decay time $\tau(k)$ obtained from the width at half maximum of the Fourier-transform of the non-equilibrium occupations $\vert n(k,\omega)\rvert$ at given k.

\begin{figure}[t]
	\centering
	\includegraphics[width=1.0\linewidth]{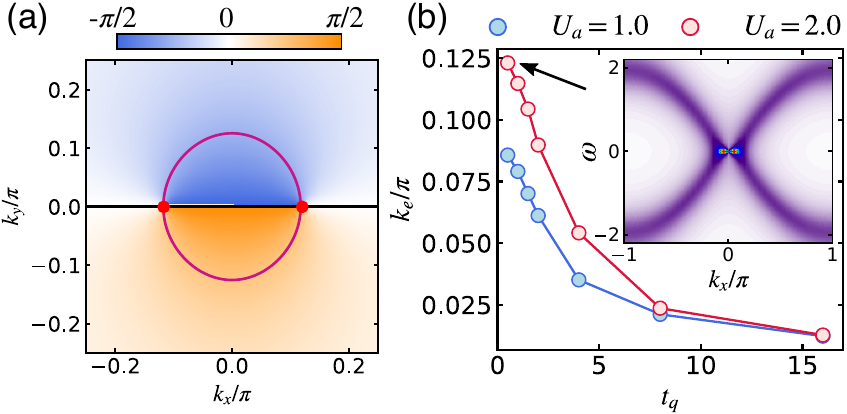}
	\caption{\label{gkba} (a) Phase of the complex energy gap function analogous to Fig.~\ref{fig_1D_1}(b) for $t_q = t_\mathrm{I}=0.5$ and $U_a=2$. The violet (black) contours represent the condition Eq.~\eqref{cond_1} (Eq.~\eqref{cond_2}), while the red dots indicate the EPs. (b) Position of the EPs as function of the quench time $t_q$. For $U_a=1$ we used the GKBA for a $N_k=96\times 96$ sampling of the Brilloin zone, while for $U_a=2$ the full KBEs were solved on a $N_k=40\times 40$ grid. $U_b=0$ is fixed in all calculations. The inset shows the spectral function $A(\mathbf{k},\omega)$ with $\mathbf{k} = (k_x,0)$. }
\end{figure}

{\it Microscopic 2D model and numerical analysis.---}
With our present NEGF approach not only quite long times but even 2D systems are amenable to microscopic study. To demonstrate the emergence of exceptional NH phases in quenched 2D systems, we now consider a model that is similarly structured to the above 1D case (cf. Eqs.~\eqref{Hssh}-\eqref{sshbloch}), but with a band-structure defined by the time-dependent Bloch Hamiltonian
$H_{\text{2D}} (\mathbf{k},t) =  \mathbf{d}_{\text{2D}}(\mathbf{k},t) \cdot \bm{\sigma}$, where $\mathbf{d}_{\text{2D}}(\mathbf{k},t) = \left(m(t)-\cos(k_x)-\cos(k_y), \sin(k_x), \sin(k_y)\right)$. The system is prepared in an insulating zero-temperature state corresponding to $m(t_i)>2$, before the band-structure is quenched to the critical point at $m(t_f)=2$, where a transition to a Chern insulator phase \cite{Haldane1988,Qi2011} takes place. 
The quench is realized by a smooth variation with duration $t_q$.
Then, a nearest neighbor interaction with different strength $U_a$ ($U_b$) within sublattice A (B) is switched on (see Eq.~\eqref{Hint}) smoothly over a short time interval $t_\mathrm{I}$. The time evolution is, as for the 1D model, determined by solving the KBEs. Due to the signifantly increased computational cost we also invoke the generalized Kadanoff-Baym ansatz (GKBA) when finer resolution in momentum space and longer times are required~\footnote{Explicit comparisons show that the GKBA provides an accurate description for interaction strength up to $U_a = 1$. Details on the method and numerical implementation can be found in refs.~\cite{Schueler2019,schuler_how_2020-1}.}.

The interacting 2D system breaks the chiral symmetry~\eqref{eqn:cs}. However, in contrast to the 1D case, EPs can occur even without a protecting symmetry, as demonstrated in Fig.~\ref{gkba}(a). Consistent with symmetry considerations, the steady state exhibits EPs at momenta $(k_x,k_y)=(\pm k_e, 0)$. Due to increased phase space, the 2D system is found to thermalize efficiently on attainable time scales (even for $U_b=0$). Hence, the position of the EPs (i.e. $k_e$) can be controlled by the quench protocol via the injected entropy. In a weak-coupling scenario, the energy is expected to be predominantly determined by $t_q$, which is confirmed in Fig.~\ref{gkba}(b).

{\it Conclusion.---} In summary, we have demonstrated the emergence of exceptional NH phases in the coherent post-quench dynamics of interacting semi-metals in both 1D and 2D. Our study reveals several genuine non-equilibrium fingerprints of NH topological phases. In particular, experimental signatures of the nodal EP structures become clearly visible in spectral functions on faster time-scales than those required for a full thermalization of the considered systems. Quite remarkably, the exceptional spectral properties manifest in the transient dynamics of momentum distribution functions that are found to exhibit a strikingly slow decay along the Fermi-arcs connecting the EPs.\\

\acknowledgments
{\it Acknowledgments.---}
We would like to thank Lorenzo Pastori for discussions. We acknowledge financial support from the German Research Foundation (DFG) through the Collaborative Research Centre SFB 1143, the Cluster of Excellence ct.qmat, the DFG Project 419241108, and the support from the U.S. Department of Energy (DOE), Office of Basic Energy Sciences, Division of Materials Sciences and Engineering, under contract DE-AC02-76SF00515. M.~S. thanks the Alexander von Humboldt Foundation for its support with a Feodor Lynen scholarship. Our numerical calculations were performed on resources at the TU Dresden Center for Information Services and High Performance Computing (ZIH) and on the Stanford Research Computing Center.


\begin{thebibliography}{45}%
	\makeatletter
	\providecommand \@ifxundefined [1]{%
		\@ifx{#1\undefined}
	}%
	\providecommand \@ifnum [1]{%
		\ifnum #1\expandafter \@firstoftwo
		\else \expandafter \@secondoftwo
		\fi
	}%
	\providecommand \@ifx [1]{%
		\ifx #1\expandafter \@firstoftwo
		\else \expandafter \@secondoftwo
		\fi
	}%
	\providecommand \natexlab [1]{#1}%
	\providecommand \enquote  [1]{``#1''}%
	\providecommand \bibnamefont  [1]{#1}%
	\providecommand \bibfnamefont [1]{#1}%
	\providecommand \citenamefont [1]{#1}%
	\providecommand \href@noop [0]{\@secondoftwo}%
	\providecommand \href [0]{\begingroup \@sanitize@url \@href}%
	\providecommand \@href[1]{\@@startlink{#1}\@@href}%
	\providecommand \@@href[1]{\endgroup#1\@@endlink}%
	\providecommand \@sanitize@url [0]{\catcode `\\12\catcode `\$12\catcode
		`\&12\catcode `\#12\catcode `\^12\catcode `\_12\catcode `\%12\relax}%
	\providecommand \@@startlink[1]{}%
	\providecommand \@@endlink[0]{}%
	\providecommand \url  [0]{\begingroup\@sanitize@url \@url }%
	\providecommand \@url [1]{\endgroup\@href {#1}{\urlprefix }}%
	\providecommand \urlprefix  [0]{URL }%
	\providecommand \Eprint [0]{\href }%
	\providecommand \doibase [0]{http://dx.doi.org/}%
	\providecommand \selectlanguage [0]{\@gobble}%
	\providecommand \bibinfo  [0]{\@secondoftwo}%
	\providecommand \bibfield  [0]{\@secondoftwo}%
	\providecommand \translation [1]{[#1]}%
	\providecommand \BibitemOpen [0]{}%
	\providecommand \bibitemStop [0]{}%
	\providecommand \bibitemNoStop [0]{.\EOS\space}%
	\providecommand \EOS [0]{\spacefactor3000\relax}%
	\providecommand \BibitemShut  [1]{\csname bibitem#1\endcsname}%
	\let\auto@bib@innerbib\@empty
	\bibitem [{\citenamefont {Landau}(1956)}]{Landau1957}%
	\BibitemOpen
	\bibfield  {author} {\bibinfo {author} {\bibfnamefont {L.D.}\ \bibnamefont
			{Landau}},\ }\bibfield  {title} {\enquote {\bibinfo {title} {The theory of a
				fermi liquid},}\ }\href
	{http://www.jetp.ac.ru/cgi-bin/e/index/e/3/6/p920?a=list} {\bibfield
		{journal} {\bibinfo  {journal} {JETP}\ }\textbf {\bibinfo {volume} {3}},\
		\bibinfo {pages} {920} (\bibinfo {year} {1956})}\BibitemShut {NoStop}%
	\bibitem [{\citenamefont {Ashcroft}\ and\ \citenamefont
		{Mermin}(1976)}]{AshcroftMermin}%
	\BibitemOpen
	\bibfield  {author} {\bibinfo {author} {\bibfnamefont {N.~W.}\ \bibnamefont
			{Ashcroft}}\ and\ \bibinfo {author} {\bibfnamefont {N.D.}\ \bibnamefont
			{Mermin}},\ }\href {https://www.cengagebrain.co.uk/shop/isbn/9780030839931}
	{\emph {\bibinfo {title} {Solid State Physics}}}\ (\bibinfo  {publisher}
	{Saunders College Publishing},\ \bibinfo {year} {1976})\BibitemShut {NoStop}%
	\bibitem [{\citenamefont {Berry}(2004)}]{Berry2004}%
	\BibitemOpen
	\bibfield  {author} {\bibinfo {author} {\bibfnamefont {M~V}\ \bibnamefont
			{Berry}},\ }\bibfield  {title} {\enquote {\bibinfo {title} {{Physics of
					Nonhermitian Degeneracies}},}\ }\href {\doibase
		10.1023/B:CJOP.0000044002.05657.04} {\bibfield  {journal} {\bibinfo
			{journal} {Czechoslovak Journal of Physics}\ }\textbf {\bibinfo {volume}
			{54}},\ \bibinfo {pages} {1039--1047} (\bibinfo {year} {2004})}\BibitemShut
	{NoStop}%
	\bibitem [{\citenamefont {{Heiss}}(2012)}]{Heiss2012}%
	\BibitemOpen
	\bibfield  {author} {\bibinfo {author} {\bibfnamefont {W.~D.}\ \bibnamefont
			{{Heiss}}},\ }\bibfield  {title} {\enquote {\bibinfo {title} {{The physics of
					exceptional points}},}\ }\href {\doibase 10.1088/1751-8113/45/44/444016}
	{\bibfield  {journal} {\bibinfo  {journal} {Journal of Physics A Mathematical
				General}\ }\textbf {\bibinfo {volume} {45}},\ \bibinfo {eid} {444016}
		(\bibinfo {year} {2012})},\ \Eprint {http://arxiv.org/abs/1210.7536}
	{arXiv:1210.7536 [quant-ph]} \BibitemShut {NoStop}%
	\bibitem [{\citenamefont {Miri}\ and\ \citenamefont
		{Al{\`u}}(2019)}]{Miri2019}%
	\BibitemOpen
	\bibfield  {author} {\bibinfo {author} {\bibfnamefont {Mohammad-Ali}\
			\bibnamefont {Miri}}\ and\ \bibinfo {author} {\bibfnamefont {Andrea}\
			\bibnamefont {Al{\`u}}},\ }\bibfield  {title} {\enquote {\bibinfo {title}
			{Exceptional points in optics and photonics},}\ }\href {\doibase
		10.1126/science.aar7709} {\bibfield  {journal} {\bibinfo  {journal}
			{Science}\ }\textbf {\bibinfo {volume} {363}} (\bibinfo {year} {2019})}\BibitemShut {NoStop}%
	\bibitem [{\citenamefont {Kozii}\ and\ \citenamefont {Fu}(2017)}]{Kozii2017}%
	\BibitemOpen
	\bibfield  {author} {\bibinfo {author} {\bibfnamefont {Vladyslav}\
			\bibnamefont {Kozii}}\ and\ \bibinfo {author} {\bibfnamefont {Liang}\
			\bibnamefont {Fu}},\ }\bibfield  {title} {\enquote {\bibinfo {title}
			{Non-hermitian topological theory of finite-lifetime quasiparticles:
				Prediction of bulk fermi arc due to exceptional point},}\ }\href@noop {} {\
		(\bibinfo {year} {2017})},\ \Eprint {http://arxiv.org/abs/1708.05841}
	{arXiv:1708.05841 [cond-mat.mes-hall]} \BibitemShut {NoStop}%
	\bibitem [{\citenamefont {Bergholtz}\ and\ \citenamefont
		{Budich}(2019)}]{Bergholtz2019}%
	\BibitemOpen
	\bibfield  {author} {\bibinfo {author} {\bibfnamefont {Emil~J.}\ \bibnamefont
			{Bergholtz}}\ and\ \bibinfo {author} {\bibfnamefont {Jan~Carl}\ \bibnamefont
			{Budich}},\ }\bibfield  {title} {\enquote {\bibinfo {title} {Non-hermitian
				weyl physics in topological insulator ferromagnet junctions},}\ }\href
	{http://dx.doi.org/10.1103/PhysRevResearch.1.012003} {\bibfield  {journal}
		{\bibinfo  {journal} {Physical Review Research}\ }\textbf {\bibinfo {volume}
			{1}} (\bibinfo {year} {2019})}\BibitemShut {NoStop}%
	\bibitem [{\citenamefont {Yoshida}\ \emph {et~al.}(2020)\citenamefont
		{Yoshida}, \citenamefont {Peters}, \citenamefont {Kawakami},\ and\
		\citenamefont {Hatsugai}}]{Yoshida2020}%
	\BibitemOpen
	\bibfield  {author} {\bibinfo {author} {\bibfnamefont {Tsuneya}\ \bibnamefont
			{Yoshida}}, \bibinfo {author} {\bibfnamefont {Robert}\ \bibnamefont
			{Peters}}, \bibinfo {author} {\bibfnamefont {Norio}\ \bibnamefont
			{Kawakami}}, \ and\ \bibinfo {author} {\bibfnamefont {Yasuhiro}\ \bibnamefont
			{Hatsugai}},\ }\bibfield  {title} {\enquote {\bibinfo {title} {Exceptional
				band touching for strongly correlated systems in equilibrium},}\ }\href
	{http://dx.doi.org/10.1093/ptep/ptaa059} {\bibfield  {journal} {\bibinfo
			{journal} {Progress of Theoretical and Experimental Physics}\ }\textbf
		{\bibinfo {volume} {2020}} (\bibinfo {year} {2020})}\BibitemShut {NoStop}%
	\bibitem [{\citenamefont {Rausch}\ \emph {et~al.}(2021)\citenamefont {Rausch},
		\citenamefont {Peters},\ and\ \citenamefont {Yoshida}}]{Rausch2021}%
	\BibitemOpen
	\bibfield  {author} {\bibinfo {author} {\bibfnamefont {Roman}\ \bibnamefont
			{Rausch}}, \bibinfo {author} {\bibfnamefont {Robert}\ \bibnamefont {Peters}},
		\ and\ \bibinfo {author} {\bibfnamefont {Tsuneya}\ \bibnamefont {Yoshida}},\
	}\bibfield  {title} {\enquote {\bibinfo {title} {Exceptional points in the
				one-dimensional hubbard model},}\ }\href {\doibase 10.1088/1367-2630/abd35e}
	{\bibfield  {journal} {\bibinfo  {journal} {New Journal of Physics}\ }\textbf
		{\bibinfo {volume} {23}},\ \bibinfo {pages} {013011} (\bibinfo {year}
		{2021})}\BibitemShut {NoStop}%
	\bibitem [{\citenamefont {Rudner}\ and\ \citenamefont
		{Levitov}(2009)}]{Rudner2009}%
	\BibitemOpen
	\bibfield  {author} {\bibinfo {author} {\bibfnamefont {M.~S.}\ \bibnamefont
			{Rudner}}\ and\ \bibinfo {author} {\bibfnamefont {L.~S.}\ \bibnamefont
			{Levitov}},\ }\bibfield  {title} {\enquote {\bibinfo {title} {Topological
				transition in a non-hermitian quantum walk},}\ }\href {\doibase
		10.1103/PhysRevLett.102.065703} {\bibfield  {journal} {\bibinfo  {journal}
			{Phys. Rev. Lett.}\ }\textbf {\bibinfo {volume} {102}},\ \bibinfo {pages}
		{065703} (\bibinfo {year} {2009})}\BibitemShut {NoStop}%
	\bibitem [{\citenamefont {Zeuner}\ \emph {et~al.}(2015)\citenamefont {Zeuner},
		\citenamefont {Rechtsman}, \citenamefont {Plotnik}, \citenamefont {Lumer},
		\citenamefont {Nolte}, \citenamefont {Rudner}, \citenamefont {Segev},\ and\
		\citenamefont {Szameit}}]{Zeuner2015}%
	\BibitemOpen
	\bibfield  {author} {\bibinfo {author} {\bibfnamefont {Julia~M.}\
			\bibnamefont {Zeuner}}, \bibinfo {author} {\bibfnamefont {Mikael~C.}\
			\bibnamefont {Rechtsman}}, \bibinfo {author} {\bibfnamefont {Yonatan}\
			\bibnamefont {Plotnik}}, \bibinfo {author} {\bibfnamefont {Yaakov}\
			\bibnamefont {Lumer}}, \bibinfo {author} {\bibfnamefont {Stefan}\
			\bibnamefont {Nolte}}, \bibinfo {author} {\bibfnamefont {Mark~S.}\
			\bibnamefont {Rudner}}, \bibinfo {author} {\bibfnamefont {Mordechai}\
			\bibnamefont {Segev}}, \ and\ \bibinfo {author} {\bibfnamefont {Alexander}\
			\bibnamefont {Szameit}},\ }\bibfield  {title} {\enquote {\bibinfo {title}
			{Observation of a topological transition in the bulk of a non-hermitian
				system},}\ }\href {\doibase 10.1103/PhysRevLett.115.040402} {\bibfield
		{journal} {\bibinfo  {journal} {Phys. Rev. Lett.}\ }\textbf {\bibinfo
			{volume} {115}},\ \bibinfo {pages} {040402} (\bibinfo {year}
		{2015})}\BibitemShut {NoStop}%
	\bibitem [{\citenamefont {Malzard}\ \emph {et~al.}(2015)\citenamefont
		{Malzard}, \citenamefont {Poli},\ and\ \citenamefont
		{Schomerus}}]{Malzard2015}%
	\BibitemOpen
	\bibfield  {author} {\bibinfo {author} {\bibfnamefont {Simon}\ \bibnamefont
			{Malzard}}, \bibinfo {author} {\bibfnamefont {Charles}\ \bibnamefont {Poli}},
		\ and\ \bibinfo {author} {\bibfnamefont {Henning}\ \bibnamefont
			{Schomerus}},\ }\bibfield  {title} {\enquote {\bibinfo {title} {Topologically
				protected defect states in open photonic systems with non-hermitian
				charge-conjugation and parity-time symmetry},}\ }\href {\doibase
		10.1103/PhysRevLett.115.200402} {\bibfield  {journal} {\bibinfo  {journal}
			{Phys. Rev. Lett.}\ }\textbf {\bibinfo {volume} {115}},\ \bibinfo {pages}
		{200402} (\bibinfo {year} {2015})}\BibitemShut {NoStop}%
	\bibitem [{\citenamefont {Lee}(2016)}]{Lee2016}%
	\BibitemOpen
	\bibfield  {author} {\bibinfo {author} {\bibfnamefont {Tony~E.}\ \bibnamefont
			{Lee}},\ }\bibfield  {title} {\enquote {\bibinfo {title} {Anomalous edge
				state in a non-hermitian lattice},}\ }\href {\doibase
		10.1103/PhysRevLett.116.133903} {\bibfield  {journal} {\bibinfo  {journal}
			{Phys. Rev. Lett.}\ }\textbf {\bibinfo {volume} {116}},\ \bibinfo {pages}
		{133903} (\bibinfo {year} {2016})}\BibitemShut {NoStop}%
	\bibitem [{\citenamefont {Zhou}\ \emph {et~al.}(2018)\citenamefont {Zhou},
		\citenamefont {Peng}, \citenamefont {Yoon}, \citenamefont {Hsu},
		\citenamefont {Nelson}, \citenamefont {Fu}, \citenamefont {Joannopoulos},
		\citenamefont {Solja{\v c}i{\'c}},\ and\ \citenamefont {Zhen}}]{Zhou2018}%
	\BibitemOpen
	\bibfield  {author} {\bibinfo {author} {\bibfnamefont {Hengyun}\ \bibnamefont
			{Zhou}}, \bibinfo {author} {\bibfnamefont {Chao}\ \bibnamefont {Peng}},
		\bibinfo {author} {\bibfnamefont {Yoseob}\ \bibnamefont {Yoon}}, \bibinfo
		{author} {\bibfnamefont {Chia~Wei}\ \bibnamefont {Hsu}}, \bibinfo {author}
		{\bibfnamefont {Keith~A.}\ \bibnamefont {Nelson}}, \bibinfo {author}
		{\bibfnamefont {Liang}\ \bibnamefont {Fu}}, \bibinfo {author} {\bibfnamefont
			{John~D.}\ \bibnamefont {Joannopoulos}}, \bibinfo {author} {\bibfnamefont
			{Marin}\ \bibnamefont {Solja{\v c}i{\'c}}}, \ and\ \bibinfo {author}
		{\bibfnamefont {Bo}~\bibnamefont {Zhen}},\ }\bibfield  {title} {\enquote
		{\bibinfo {title} {Observation of bulk fermi arc and polarization half charge
				from paired exceptional points},}\ }\href {\doibase 10.1126/science.aap9859}
	{\bibfield  {journal} {\bibinfo  {journal} {Science}\ }\textbf {\bibinfo
			{volume} {359}},\ \bibinfo {pages} {1009--1012} (\bibinfo {year}
		{2018})}\BibitemShut {NoStop}%
	\bibitem [{\citenamefont {Lieu}(2018)}]{Lieu2018}%
	\BibitemOpen
	\bibfield  {author} {\bibinfo {author} {\bibfnamefont {Simon}\ \bibnamefont
			{Lieu}},\ }\bibfield  {title} {\enquote {\bibinfo {title} {Topological phases
				in the non-hermitian su-schrieffer-heeger model},}\ }\href {\doibase
		10.1103/PhysRevB.97.045106} {\bibfield  {journal} {\bibinfo  {journal} {Phys.
				Rev. B}\ }\textbf {\bibinfo {volume} {97}},\ \bibinfo {pages} {045106}
		(\bibinfo {year} {2018})}\BibitemShut {NoStop}%
	\bibitem [{\citenamefont {Longhi}(2018)}]{Longhi2018}%
	\BibitemOpen
	\bibfield  {author} {\bibinfo {author} {\bibfnamefont {Stefano}\ \bibnamefont
			{Longhi}},\ }\bibfield  {title} {\enquote {\bibinfo {title} {Non-hermitian
				gauged topological laser arrays},}\ }\href {\doibase
		https://doi.org/10.1002/andp.201800023} {\bibfield  {journal} {\bibinfo
			{journal} {Annalen der Physik}\ }\textbf {\bibinfo {volume} {530}},\ \bibinfo
		{pages} {1800023} (\bibinfo {year} {2018})}\BibitemShut {NoStop}%
	\bibitem [{\citenamefont {Kunst}\ \emph {et~al.}(2018)\citenamefont {Kunst},
		\citenamefont {Edvardsson}, \citenamefont {Budich},\ and\ \citenamefont
		{Bergholtz}}]{Kunst2018}%
	\BibitemOpen
	\bibfield  {author} {\bibinfo {author} {\bibfnamefont {Flore~K.}\
			\bibnamefont {Kunst}}, \bibinfo {author} {\bibfnamefont {Elisabet}\
			\bibnamefont {Edvardsson}}, \bibinfo {author} {\bibfnamefont {Jan~Carl}\
			\bibnamefont {Budich}}, \ and\ \bibinfo {author} {\bibfnamefont {Emil~J.}\
			\bibnamefont {Bergholtz}},\ }\bibfield  {title} {\enquote {\bibinfo {title}
			{Biorthogonal bulk-boundary correspondence in non-hermitian systems},}\
	}\href {\doibase 10.1103/PhysRevLett.121.026808} {\bibfield  {journal}
		{\bibinfo  {journal} {Phys. Rev. Lett.}\ }\textbf {\bibinfo {volume} {121}},\
		\bibinfo {pages} {026808} (\bibinfo {year} {2018})}\BibitemShut {NoStop}%
	\bibitem [{\citenamefont {Yao}\ and\ \citenamefont {Wang}(2018)}]{Yao2018}%
	\BibitemOpen
	\bibfield  {author} {\bibinfo {author} {\bibfnamefont {Shunyu}\ \bibnamefont
			{Yao}}\ and\ \bibinfo {author} {\bibfnamefont {Zhong}\ \bibnamefont {Wang}},\
	}\bibfield  {title} {\enquote {\bibinfo {title} {Edge states and topological
				invariants of non-hermitian systems},}\ }\href {\doibase
		10.1103/PhysRevLett.121.086803} {\bibfield  {journal} {\bibinfo  {journal}
			{Phys. Rev. Lett.}\ }\textbf {\bibinfo {volume} {121}},\ \bibinfo {pages}
		{086803} (\bibinfo {year} {2018})}\BibitemShut {NoStop}%
	\bibitem [{\citenamefont {Imhof}\ \emph {et~al.}(2018)\citenamefont {Imhof},
		\citenamefont {Berger},\ and\ \citenamefont {Bayer}}]{Imhof2018}%
	\BibitemOpen
	\bibfield  {author} {\bibinfo {author} {\bibfnamefont {Stefan}\ \bibnamefont
			{Imhof}}, \bibinfo {author} {\bibfnamefont {Christian}\ \bibnamefont
			{Berger}}, \ and\ \bibinfo {author} {\bibfnamefont {Florian et~al.}\
			\bibnamefont {Bayer}},\ }\bibfield  {title} {\enquote {\bibinfo {title}
			{Topolectrical-circuit realization of topological corner modes},}\ }\href
	{\doibase 10.1038/s41567-018-0246-1} {\bibfield  {journal} {\bibinfo
			{journal} {Nature Physics}\ }\textbf {\bibinfo {volume} {14}},\ \bibinfo
		{pages} {925--929} (\bibinfo {year} {2018})}\BibitemShut {NoStop}%
	\bibitem [{\citenamefont {Gong}\ \emph {et~al.}(2018)\citenamefont {Gong},
		\citenamefont {Ashida}, \citenamefont {Kawabata}, \citenamefont {Takasan},
		\citenamefont {Higashikawa},\ and\ \citenamefont {Ueda}}]{Gong2018}%
	\BibitemOpen
	\bibfield  {author} {\bibinfo {author} {\bibfnamefont {Zongping}\
			\bibnamefont {Gong}}, \bibinfo {author} {\bibfnamefont {Yuto}\ \bibnamefont
			{Ashida}}, \bibinfo {author} {\bibfnamefont {Kohei}\ \bibnamefont
			{Kawabata}}, \bibinfo {author} {\bibfnamefont {Kazuaki}\ \bibnamefont
			{Takasan}}, \bibinfo {author} {\bibfnamefont {Sho}\ \bibnamefont
			{Higashikawa}}, \ and\ \bibinfo {author} {\bibfnamefont {Masahito}\
			\bibnamefont {Ueda}},\ }\bibfield  {title} {\enquote {\bibinfo {title}
			{Topological phases of non-hermitian systems},}\ }\href {\doibase
		10.1103/PhysRevX.8.031079} {\bibfield  {journal} {\bibinfo  {journal} {Phys.
				Rev. X}\ }\textbf {\bibinfo {volume} {8}},\ \bibinfo {pages} {031079}
		(\bibinfo {year} {2018})}\BibitemShut {NoStop}%
	\bibitem [{\citenamefont {Kawabata}\ \emph
		{et~al.}(2019{\natexlab{a}})\citenamefont {Kawabata}, \citenamefont
		{Bessho},\ and\ \citenamefont {Sato}}]{Kawabata2019_1}%
	\BibitemOpen
	\bibfield  {author} {\bibinfo {author} {\bibfnamefont {Kohei}\ \bibnamefont
			{Kawabata}}, \bibinfo {author} {\bibfnamefont {Takumi}\ \bibnamefont
			{Bessho}}, \ and\ \bibinfo {author} {\bibfnamefont {Masatoshi}\ \bibnamefont
			{Sato}},\ }\bibfield  {title} {\enquote {\bibinfo {title} {Classification of
				exceptional points and non-hermitian topological semimetals},}\ }\href
	{\doibase 10.1103/PhysRevLett.123.066405} {\bibfield  {journal} {\bibinfo
			{journal} {Phys. Rev. Lett.}\ }\textbf {\bibinfo {volume} {123}},\ \bibinfo
		{pages} {066405} (\bibinfo {year} {2019}{\natexlab{a}})}\BibitemShut
	{NoStop}%
	\bibitem [{\citenamefont {Budich}\ \emph {et~al.}(2019)\citenamefont {Budich},
		\citenamefont {Carlstr\"om}, \citenamefont {Kunst},\ and\ \citenamefont
		{Bergholtz}}]{Budich2019}%
	\BibitemOpen
	\bibfield  {author} {\bibinfo {author} {\bibfnamefont {Jan~Carl}\
			\bibnamefont {Budich}}, \bibinfo {author} {\bibfnamefont {Johan}\
			\bibnamefont {Carlstr\"om}}, \bibinfo {author} {\bibfnamefont {Flore~K.}\
			\bibnamefont {Kunst}}, \ and\ \bibinfo {author} {\bibfnamefont {Emil~J.}\
			\bibnamefont {Bergholtz}},\ }\bibfield  {title} {\enquote {\bibinfo {title}
			{Symmetry-protected nodal phases in non-hermitian systems},}\ }\href
	{\doibase 10.1103/PhysRevB.99.041406} {\bibfield  {journal} {\bibinfo
			{journal} {Phys. Rev. B}\ }\textbf {\bibinfo {volume} {99}},\ \bibinfo
		{pages} {041406} (\bibinfo {year} {2019})}\BibitemShut {NoStop}%
	\bibitem [{\citenamefont {Yoshida}\ \emph {et~al.}(2019)\citenamefont
		{Yoshida}, \citenamefont {Peters}, \citenamefont {Kawakami},\ and\
		\citenamefont {Hatsugai}}]{Yoshida2019}%
	\BibitemOpen
	\bibfield  {author} {\bibinfo {author} {\bibfnamefont {Tsuneya}\ \bibnamefont
			{Yoshida}}, \bibinfo {author} {\bibfnamefont {Robert}\ \bibnamefont
			{Peters}}, \bibinfo {author} {\bibfnamefont {Norio}\ \bibnamefont
			{Kawakami}}, \ and\ \bibinfo {author} {\bibfnamefont {Yasuhiro}\ \bibnamefont
			{Hatsugai}},\ }\bibfield  {title} {\enquote {\bibinfo {title}
			{Symmetry-protected exceptional rings in two-dimensional correlated systems
				with chiral symmetry},}\ }\href {\doibase 10.1103/PhysRevB.99.121101}
	{\bibfield  {journal} {\bibinfo  {journal} {Phys. Rev. B}\ }\textbf {\bibinfo
			{volume} {99}},\ \bibinfo {pages} {121101} (\bibinfo {year}
		{2019})}\BibitemShut {NoStop}%
	\bibitem [{\citenamefont {Cerjan}\ \emph {et~al.}(2019)\citenamefont {Cerjan},
		\citenamefont {Huang},\ and\ \citenamefont {Wang}}]{Cerjan2019}%
	\BibitemOpen
	\bibfield  {author} {\bibinfo {author} {\bibfnamefont {Alexander}\
			\bibnamefont {Cerjan}}, \bibinfo {author} {\bibfnamefont {Sheng}\
			\bibnamefont {Huang}}, \ and\ \bibinfo {author} {\bibfnamefont {Mohan
				et~al.}\ \bibnamefont {Wang}},\ }\bibfield  {title} {\enquote {\bibinfo
			{title} {Experimental realization of a weyl exceptional ring},}\ }\href
	{\doibase 10.1038/s41566-019-0453-z} {\bibfield  {journal} {\bibinfo
			{journal} {Nature Physics}\ }\textbf {\bibinfo {volume} {13}},\ \bibinfo
		{pages} {623--628} (\bibinfo {year} {2019})}\BibitemShut {NoStop}%
	\bibitem [{\citenamefont {Kawabata}\ \emph
		{et~al.}(2019{\natexlab{b}})\citenamefont {Kawabata}, \citenamefont
		{Shiozaki}, \citenamefont {Ueda},\ and\ \citenamefont
		{Sato}}]{Kawabata2019_2}%
	\BibitemOpen
	\bibfield  {author} {\bibinfo {author} {\bibfnamefont {Kohei}\ \bibnamefont
			{Kawabata}}, \bibinfo {author} {\bibfnamefont {Ken}\ \bibnamefont
			{Shiozaki}}, \bibinfo {author} {\bibfnamefont {Masahito}\ \bibnamefont
			{Ueda}}, \ and\ \bibinfo {author} {\bibfnamefont {Masatoshi}\ \bibnamefont
			{Sato}},\ }\bibfield  {title} {\enquote {\bibinfo {title} {Symmetry and
				topology in non-hermitian physics},}\ }\href {\doibase
		10.1103/PhysRevX.9.041015} {\bibfield  {journal} {\bibinfo  {journal} {Phys.
				Rev. X}\ }\textbf {\bibinfo {volume} {9}},\ \bibinfo {pages} {041015}
		(\bibinfo {year} {2019}{\natexlab{b}})}\BibitemShut {NoStop}%
	\bibitem [{\citenamefont {Kunst}\ and\ \citenamefont
		{Dwivedi}(2019)}]{Kunst2019}%
	\BibitemOpen
	\bibfield  {author} {\bibinfo {author} {\bibfnamefont {Flore~K.}\
			\bibnamefont {Kunst}}\ and\ \bibinfo {author} {\bibfnamefont {Vatsal}\
			\bibnamefont {Dwivedi}},\ }\bibfield  {title} {\enquote {\bibinfo {title}
			{Non-hermitian systems and topology: A transfer-matrix perspective},}\ }\href
	{\doibase 10.1103/PhysRevB.99.245116} {\bibfield  {journal} {\bibinfo
			{journal} {Phys. Rev. B}\ }\textbf {\bibinfo {volume} {99}},\ \bibinfo
		{pages} {245116} (\bibinfo {year} {2019})}\BibitemShut {NoStop}%
	\bibitem [{\citenamefont {Song}\ \emph {et~al.}(2019)\citenamefont {Song},
		\citenamefont {Yao},\ and\ \citenamefont {Wang}}]{Song2019}%
	\BibitemOpen
	\bibfield  {author} {\bibinfo {author} {\bibfnamefont {Fei}\ \bibnamefont
			{Song}}, \bibinfo {author} {\bibfnamefont {Shunyu}\ \bibnamefont {Yao}}, \
		and\ \bibinfo {author} {\bibfnamefont {Zhong}\ \bibnamefont {Wang}},\
	}\bibfield  {title} {\enquote {\bibinfo {title} {Non-hermitian skin effect
				and chiral damping in open quantum systems},}\ }\href {\doibase
		10.1103/PhysRevLett.123.170401} {\bibfield  {journal} {\bibinfo  {journal}
			{Phys. Rev. Lett.}\ }\textbf {\bibinfo {volume} {123}},\ \bibinfo {pages}
		{170401} (\bibinfo {year} {2019})}\BibitemShut {NoStop}%
	\bibitem [{\citenamefont {Xiao}\ \emph {et~al.}(2020)\citenamefont {Xiao},
		\citenamefont {Deng},\ and\ \citenamefont {Wang}}]{Xiao2020}%
	\BibitemOpen
	\bibfield  {author} {\bibinfo {author} {\bibfnamefont {Lei}\ \bibnamefont
			{Xiao}}, \bibinfo {author} {\bibfnamefont {Tianshu}\ \bibnamefont {Deng}}, \
		and\ \bibinfo {author} {\bibfnamefont {Kunkun et~al.}\ \bibnamefont {Wang}},\
	}\bibfield  {title} {\enquote {\bibinfo {title} {Non-hermitian
				bulk–boundary correspondence in quantum dynamics},}\ }\href {\doibase
		10.1038/s41567-020-0836-6} {\bibfield  {journal} {\bibinfo  {journal} {Nature
				Physics}\ }\textbf {\bibinfo {volume} {16}},\ \bibinfo {pages} {761--766}
		(\bibinfo {year} {2020})}\BibitemShut {NoStop}%
	\bibitem [{\citenamefont {Weidemann}\ \emph {et~al.}(2020)\citenamefont
		{Weidemann}, \citenamefont {Kremer}, \citenamefont {Helbig}, \citenamefont
		{Hofmann}, \citenamefont {Stegmaier}, \citenamefont {Greiter}, \citenamefont
		{Thomale},\ and\ \citenamefont {Szameit}}]{Weidemann2020}%
	\BibitemOpen
	\bibfield  {author} {\bibinfo {author} {\bibfnamefont {Sebastian}\
			\bibnamefont {Weidemann}}, \bibinfo {author} {\bibfnamefont {Mark}\
			\bibnamefont {Kremer}}, \bibinfo {author} {\bibfnamefont {Tobias}\
			\bibnamefont {Helbig}}, \bibinfo {author} {\bibfnamefont {Tobias}\
			\bibnamefont {Hofmann}}, \bibinfo {author} {\bibfnamefont {Alexander}\
			\bibnamefont {Stegmaier}}, \bibinfo {author} {\bibfnamefont {Martin}\
			\bibnamefont {Greiter}}, \bibinfo {author} {\bibfnamefont {Ronny}\
			\bibnamefont {Thomale}}, \ and\ \bibinfo {author} {\bibfnamefont {Alexander}\
			\bibnamefont {Szameit}},\ }\bibfield  {title} {\enquote {\bibinfo {title}
			{Topological funneling of light},}\ }\href {\doibase 10.1126/science.aaz8727}
	{\bibfield  {journal} {\bibinfo  {journal} {Science}\ }\textbf {\bibinfo
			{volume} {368}},\ \bibinfo {pages} {311--314} (\bibinfo {year}
		{2020})}\BibitemShut {NoStop}%
	\bibitem [{\citenamefont {Budich}\ and\ \citenamefont
		{Bergholtz}(2020)}]{Budich2020}%
	\BibitemOpen
	\bibfield  {author} {\bibinfo {author} {\bibfnamefont {Jan~Carl}\
			\bibnamefont {Budich}}\ and\ \bibinfo {author} {\bibfnamefont {Emil~J.}\
			\bibnamefont {Bergholtz}},\ }\bibfield  {title} {\enquote {\bibinfo {title}
			{Non-hermitian topological sensors},}\ }\href {\doibase
		10.1103/PhysRevLett.125.180403} {\bibfield  {journal} {\bibinfo  {journal}
			{Phys. Rev. Lett.}\ }\textbf {\bibinfo {volume} {125}},\ \bibinfo {pages}
		{180403} (\bibinfo {year} {2020})}\BibitemShut {NoStop}%
	\bibitem [{\citenamefont {Bergholtz}\ \emph {et~al.}(2021)\citenamefont
		{Bergholtz}, \citenamefont {Budich},\ and\ \citenamefont {Kunst}}]{review}%
	\BibitemOpen
	\bibfield  {author} {\bibinfo {author} {\bibfnamefont {Emil~J.}\ \bibnamefont
			{Bergholtz}}, \bibinfo {author} {\bibfnamefont {Jan~Carl}\ \bibnamefont
			{Budich}}, \ and\ \bibinfo {author} {\bibfnamefont {Flore~K.}\ \bibnamefont
			{Kunst}},\ }\bibfield  {title} {\enquote {\bibinfo {title} {Exceptional
				topology of non-hermitian systems},}\ }\href {\doibase
		10.1103/RevModPhys.93.015005} {\bibfield  {journal} {\bibinfo  {journal}
			{Rev. Mod. Phys.}\ }\textbf {\bibinfo {volume} {93}},\ \bibinfo {pages}
		{015005} (\bibinfo {year} {2021})}\BibitemShut {NoStop}%
	\bibitem [{\citenamefont {Stefanucci}\ and\ \citenamefont {van
			Leeuwen}(2013)}]{Stefanucci_book}%
	\BibitemOpen
	\bibfield  {author} {\bibinfo {author} {\bibfnamefont {Gianluca}\
			\bibnamefont {Stefanucci}}\ and\ \bibinfo {author} {\bibfnamefont {Robert}\
			\bibnamefont {van Leeuwen}},\ }\href {\doibase 10.1017/CBO9781139023979}
	{\emph {\bibinfo {title} {Nonequilibrium Many-Body Theory of Quantum Systems:
				A Modern Introduction}}}\ (\bibinfo  {publisher} {Cambridge University
		Press},\ \bibinfo {year} {2013})\BibitemShut {NoStop}%
	\bibitem [{\citenamefont {Balzer}\ and\ \citenamefont
		{Bonitz}(2012)}]{balzer_nonequilibrium_2012}%
	\BibitemOpen
	\bibfield  {author} {\bibinfo {author} {\bibfnamefont {Karsten}\ \bibnamefont
			{Balzer}}\ and\ \bibinfo {author} {\bibfnamefont {Michael}\ \bibnamefont
			{Bonitz}},\ }\href@noop {} {\emph {\bibinfo {title} {Nonequilibrium {Green}'s
				{Functions} {Approach} to {Inhomogeneous} {Systems}}}}\ (\bibinfo
	{publisher} {Springer},\ \bibinfo {year} {2012})\BibitemShut {NoStop}%
	\bibitem [{\citenamefont {Aoki}\ \emph {et~al.}(2014)\citenamefont {Aoki},
		\citenamefont {Tsuji}, \citenamefont {Eckstein}, \citenamefont {Kollar},
		\citenamefont {Oka},\ and\ \citenamefont
		{Werner}}]{aoki_nonequilibrium_2014}%
	\BibitemOpen
	\bibfield  {author} {\bibinfo {author} {\bibfnamefont {Hideo}\ \bibnamefont
			{Aoki}}, \bibinfo {author} {\bibfnamefont {Naoto}\ \bibnamefont {Tsuji}},
		\bibinfo {author} {\bibfnamefont {Martin}\ \bibnamefont {Eckstein}}, \bibinfo
		{author} {\bibfnamefont {Marcus}\ \bibnamefont {Kollar}}, \bibinfo {author}
		{\bibfnamefont {Takashi}\ \bibnamefont {Oka}}, \ and\ \bibinfo {author}
		{\bibfnamefont {Philipp}\ \bibnamefont {Werner}},\ }\bibfield  {title}
	{\enquote {\bibinfo {title} {Nonequilibrium dynamical mean-field theory and
				its applications},}\ }\href
	{https://link.aps.org/doi/10.1103/RevModPhys.86.779} {\bibfield  {journal}
		{\bibinfo  {journal} {Rev. Mod. Phys.}\ }\textbf {\bibinfo {volume} {86}},\
		\bibinfo {pages} {779--837} (\bibinfo {year} {2014})}\BibitemShut {NoStop}%
	\bibitem [{\citenamefont {Schl\"{u}nzen}\ and\ \citenamefont
		{Bonitz}(2016)}]{schlunzen_nonequilibrium_2016}%
	\BibitemOpen
	\bibfield  {author} {\bibinfo {author} {\bibfnamefont {N.}~\bibnamefont
			{Schl\"{u}nzen}}\ and\ \bibinfo {author} {\bibfnamefont {M.}~\bibnamefont
			{Bonitz}},\ }\bibfield  {title} {\enquote {\bibinfo {title} {Nonequilibrium
				{Green} {Functions} {Approach} to {Strongly} {Correlated} {Fermions} in
				{Lattice} {Systems}},}\ }\href
	{http://onlinelibrary.wiley.com/doi/10.1002/ctpp.201610003/abstract}
	{\bibfield  {journal} {\bibinfo  {journal} {Contrib. Plasma Phys.}\ }\textbf
		{\bibinfo {volume} {56}},\ \bibinfo {pages} {5--91} (\bibinfo {year}
		{2016})}\BibitemShut {NoStop}%
	\bibitem [{\citenamefont {Sch\"uler}\ \emph {et~al.}(2020)\citenamefont
		{Sch\"uler}, \citenamefont {Golež}, \citenamefont {Murakami}, \citenamefont
		{Bittner}, \citenamefont {Herrmann}, \citenamefont {Strand}, \citenamefont
		{Werner},\ and\ \citenamefont {Eckstein}}]{nessi2020}%
	\BibitemOpen
	\bibfield  {author} {\bibinfo {author} {\bibfnamefont {Michael}\ \bibnamefont
			{Sch\"uler}}, \bibinfo {author} {\bibfnamefont {Denis}\ \bibnamefont
			{Golež}}, \bibinfo {author} {\bibfnamefont {Yuta}\ \bibnamefont {Murakami}},
		\bibinfo {author} {\bibfnamefont {Nikolaj}\ \bibnamefont {Bittner}}, \bibinfo
		{author} {\bibfnamefont {Andreas}\ \bibnamefont {Herrmann}}, \bibinfo
		{author} {\bibfnamefont {Hugo~U.R.}\ \bibnamefont {Strand}}, \bibinfo
		{author} {\bibfnamefont {Philipp}\ \bibnamefont {Werner}}, \ and\ \bibinfo
		{author} {\bibfnamefont {Martin}\ \bibnamefont {Eckstein}},\ }\bibfield
	{title} {\enquote {\bibinfo {title} {Nessi: The non-equilibrium systems
				simulation package},}\ }\href {\doibase
		https://doi.org/10.1016/j.cpc.2020.107484} {\bibfield  {journal} {\bibinfo
			{journal} {Computer Physics Communications}\ }\textbf {\bibinfo {volume}
			{257}},\ \bibinfo {pages} {107484} (\bibinfo {year} {2020})}\BibitemShut
	{NoStop}%
	\bibitem [{Note1()}]{Note1}%
	\BibitemOpen
	\bibinfo {note} {We calculated the effective Hamiltonian $H_{\protect \text
			{NH}}(k,\omega )$ by Eq.~(\ref {Kadanoff_Baym}) with the retarded Green's
		function and by Eq.~({\ref {eqn:heff}}) with the retarded self-energy $\Sigma
		^R(k,\omega )$. Both ways lead to the same results, in this paper we used the
		latter method, because $\Sigma ^R(k,\Delta t)$ decays much faster in the late
		time regime, thus Fourier transformation was numerically easier.}\BibitemShut
	{Stop}%
	\bibitem [{\citenamefont {Su}\ \emph {et~al.}(1979)\citenamefont {Su},
		\citenamefont {Schrieffer},\ and\ \citenamefont {Heeger}}]{Su1979}%
	\BibitemOpen
	\bibfield  {author} {\bibinfo {author} {\bibfnamefont {W.~P.}\ \bibnamefont
			{Su}}, \bibinfo {author} {\bibfnamefont {J.~R.}\ \bibnamefont {Schrieffer}},
		\ and\ \bibinfo {author} {\bibfnamefont {A.~J.}\ \bibnamefont {Heeger}},\
	}\bibfield  {title} {\enquote {\bibinfo {title} {Solitons in
				polyacetylene},}\ }\href {\doibase 10.1103/PhysRevLett.42.1698} {\bibfield
		{journal} {\bibinfo  {journal} {Phys. Rev. Lett.}\ }\textbf {\bibinfo
			{volume} {42}},\ \bibinfo {pages} {1698--1701} (\bibinfo {year}
		{1979})}\BibitemShut {NoStop}%
	\bibitem [{\citenamefont {Kruckenhauser}\ and\ \citenamefont
		{Budich}(2018)}]{Kruckenhauser2018}%
	\BibitemOpen
	\bibfield  {author} {\bibinfo {author} {\bibfnamefont {Andreas}\ \bibnamefont
			{Kruckenhauser}}\ and\ \bibinfo {author} {\bibfnamefont {Jan~Carl}\
			\bibnamefont {Budich}},\ }\bibfield  {title} {\enquote {\bibinfo {title}
			{Dynamical equilibration of topological properties},}\ }\href {\doibase
		10.1103/PhysRevB.98.195124} {\bibfield  {journal} {\bibinfo  {journal} {Phys.
				Rev. B}\ }\textbf {\bibinfo {volume} {98}},\ \bibinfo {pages} {195124}
		(\bibinfo {year} {2018})}\BibitemShut {NoStop}%
	\bibitem [{\citenamefont {Chiu}\ \emph {et~al.}(2016)\citenamefont {Chiu},
		\citenamefont {Teo}, \citenamefont {Schnyder},\ and\ \citenamefont
		{Ryu}}]{Chiu2016}%
	\BibitemOpen
	\bibfield  {author} {\bibinfo {author} {\bibfnamefont {Ching-Kai}\
			\bibnamefont {Chiu}}, \bibinfo {author} {\bibfnamefont {Jeffrey C.~Y.}\
			\bibnamefont {Teo}}, \bibinfo {author} {\bibfnamefont {Andreas~P.}\
			\bibnamefont {Schnyder}}, \ and\ \bibinfo {author} {\bibfnamefont {Shinsei}\
			\bibnamefont {Ryu}},\ }\bibfield  {title} {\enquote {\bibinfo {title}
			{Classification of topological quantum matter with symmetries},}\ }\href
	{\doibase 10.1103/RevModPhys.88.035005} {\bibfield  {journal} {\bibinfo
			{journal} {Rev. Mod. Phys.}\ }\textbf {\bibinfo {volume} {88}},\ \bibinfo
		{pages} {035005} (\bibinfo {year} {2016})}\BibitemShut {NoStop}%
	\bibitem [{\citenamefont {Haldane}(1988)}]{Haldane1988}%
	\BibitemOpen
	\bibfield  {author} {\bibinfo {author} {\bibfnamefont {F.~D.~M.}\
			\bibnamefont {Haldane}},\ }\bibfield  {title} {\enquote {\bibinfo {title}
			{Model for a quantum hall effect without landau levels: Condensed-matter
				realization of the "parity anomaly"},}\ }\href {\doibase
		10.1103/PhysRevLett.61.2015} {\bibfield  {journal} {\bibinfo  {journal}
			{Phys. Rev. Lett.}\ }\textbf {\bibinfo {volume} {61}},\ \bibinfo {pages}
		{2015--2018} (\bibinfo {year} {1988})}\BibitemShut {NoStop}%
	\bibitem [{\citenamefont {Qi}\ and\ \citenamefont {Zhang}(2011)}]{Qi2011}%
	\BibitemOpen
	\bibfield  {author} {\bibinfo {author} {\bibfnamefont {Xiao-Liang}\
			\bibnamefont {Qi}}\ and\ \bibinfo {author} {\bibfnamefont {Shou-Cheng}\
			\bibnamefont {Zhang}},\ }\bibfield  {title} {\enquote {\bibinfo {title}
			{Topological insulators and superconductors},}\ }\href {\doibase
		10.1103/RevModPhys.83.1057} {\bibfield  {journal} {\bibinfo  {journal} {Rev.
				Mod. Phys.}\ }\textbf {\bibinfo {volume} {83}},\ \bibinfo {pages}
		{1057--1110} (\bibinfo {year} {2011})}\BibitemShut {NoStop}%
	\bibitem [{Note2()}]{Note2}%
	\BibitemOpen
	\bibinfo {note} {Explicit comparisons show that the GKBA provides an accurate
		description for interaction strength up to $U_a = 1$. Details on the method
		and numerical implementation can be found in refs.~\cite
		{Schueler2019,schuler_how_2020-1}.}\BibitemShut {Stop}%
	\bibitem [{\citenamefont {Sch\"uler}\ \emph {et~al.}(2019)\citenamefont
		{Sch\"uler}, \citenamefont {Budich},\ and\ \citenamefont
		{Werner}}]{Schueler2019}%
	\BibitemOpen
	\bibfield  {author} {\bibinfo {author} {\bibfnamefont {Michael}\ \bibnamefont
			{Sch\"uler}}, \bibinfo {author} {\bibfnamefont {Jan~Carl}\ \bibnamefont
			{Budich}}, \ and\ \bibinfo {author} {\bibfnamefont {Philipp}\ \bibnamefont
			{Werner}},\ }\bibfield  {title} {\enquote {\bibinfo {title} {Quench dynamics
				and hall response of interacting chern insulators},}\ }\href {\doibase
		10.1103/PhysRevB.100.041101} {\bibfield  {journal} {\bibinfo  {journal}
			{Phys. Rev. B}\ }\textbf {\bibinfo {volume} {100}},\ \bibinfo {pages}
		{041101} (\bibinfo {year} {2019})}\BibitemShut {NoStop}%
	\bibitem [{\citenamefont {Sch\"{u}ler}\ \emph {et~al.}(2020)\citenamefont
		{Sch\"{u}ler}, \citenamefont {De~Giovannini}, \citenamefont {H\"{u}bener},
		\citenamefont {Rubio}, \citenamefont {Sentef}, \citenamefont {Devereaux},\
		and\ \citenamefont {Werner}}]{schuler_how_2020-1}%
	\BibitemOpen
	\bibfield  {author} {\bibinfo {author} {\bibfnamefont {Michael}\ \bibnamefont
			{Sch\"{u}ler}}, \bibinfo {author} {\bibfnamefont {Umberto}\ \bibnamefont
			{De~Giovannini}}, \bibinfo {author} {\bibfnamefont {Hannes}\ \bibnamefont
			{H\"{u}bener}}, \bibinfo {author} {\bibfnamefont {Angel}\ \bibnamefont
			{Rubio}}, \bibinfo {author} {\bibfnamefont {Michael~A.}\ \bibnamefont
			{Sentef}}, \bibinfo {author} {\bibfnamefont {Thomas~P.}\ \bibnamefont
			{Devereaux}}, \ and\ \bibinfo {author} {\bibfnamefont {Philipp}\ \bibnamefont
			{Werner}},\ }\bibfield  {title} {\enquote {\bibinfo {title} {How {Circular}
				{Dichroism} in {Time}- and {Angle}-{Resolved} {Photoemission} {Can} {Be}
				{Used} to {Spectroscopically} {Detect} {Transient} {Topological} {States} in
				{Graphene}},}\ }\href {https://link.aps.org/doi/10.1103/PhysRevX.10.041013}
	{\bibfield  {journal} {\bibinfo  {journal} {Phys. Rev. X}\ }\textbf {\bibinfo
			{volume} {10}},\ \bibinfo {pages} {041013} (\bibinfo {year}
		{2020})}\BibitemShut {NoStop}%
\end{thebibliography}

%

\end{document}